\documentclass{natureB}

\title{A matterless double slit\\[-3ex]}

\author{Ben King$^1$, Antonino Di Piazza$^1$ and Christoph H. Keitel$^1$}

\usepackage{amsfonts}
\usepackage{amsmath}
\usepackage{amssymb}
\usepackage{revsymb}
\usepackage{graphicx}
\usepackage{mathrsfs}
\usepackage{bm}
\newcommand{\be}{\begin{equation}}
\newcommand{\ee}{\end{equation}}
\newcommand{\ba}{\begin{array}}
\newcommand{\ea}{\end{array}}
\newcommand{\bea}{\begin{eqnarray}} 
\newcommand{\eea}{\end{eqnarray}} 
\newcommand{\bd}{\begin{displaymath}}
\newcommand{\ed}{\end{displaymath}}
\newcommand{\units}[1]{\,\textrm{#1}}
\newcommand{\eps}{\varepsilon}
\newcommand{\mbf}[1]{\mathbf{#1}}
\newcommand{\trm}[1]{\textrm{#1}}

\long\def\symbolfootnote[#1]#2{\begingroup%
\def\thefootnote{\fnsymbol{footnote}}\footnote[#1]{#2}\endgroup}

\newcommand{\eqnref}[1]{Eq. (\ref{#1})}

\begin{document}

\maketitle\\[-2ex]
{\small $^{1}$ Max-Planck-Institut f\"ur Kernphysik, Saupfercheckweg 1, D-69117 Heidelberg, Germany}\\[-3ex]

\begin{abstract}
Double-slits provide incoming photons with a choice. Those that survive the passage have chosen from two possible paths which interfere to distribute them in a wave-like manner. Such wave-particle duality\cite{debroglie23} continues to be challenged\cite{scully91, wiseman95, paulus05, kiffner06} and investigated in a broad range of disciplines with electrons\cite{joennson61}, neutrons\cite{dslitneutrons88}, helium atoms\cite{carnal91}, $C_{60}$ fullerenes\cite{dslitc60}, Bose-Einstein condensa-tes\cite{ketterle97} and biological molecules\cite{dslitbio}. All variants have hitherto involved material constituents. We present a matterless double-slit scenario in which photons generated from virtual electron-positron pair annihilation in head-on collisions of a probe laser field with two ultra-intense laser beams form a double-slit interference pattern. Such electromagnetic fields are predicted to induce material-like behaviour in the vacuum, supporting elastic scattering between photons\cite{dipiazza_PRL_06, lundstroem_PRL_06}. Our double-slit scenario presents on the one hand a realisable method to observe photon-photon scattering, and demonstrates on the other, the possibility of both controlling light with light and non-locally investigating features of the quantum vacuum's structure.
\end{abstract}

According to both special relativity and Heisenberg's uncertainty principle, virtual electron-positron pairs spontaneously pop into and out of existence in vacuum, on a time scale too short to leave a trace. However, it is the polarisation of these pairs under an applied electromagnetic field which is predicted to provide a rich variety of non-linear processes\cite{heisenberg_euler36}. A fundamental scale for such vacuum polarisation effects is set by the critical field of quantum electrodynamics $E_{cr}=\sqrt{4\pi}m^2c^3/\hbar e=1.3\times 10^{16}\;\text{Vcm}^{-1}$, for electron mass $m$ and absolute charge $e$ (in our units the fine-structure constant reads $\alpha=e^2/4\pi\hbar c\approx 1/137$), corresponding to a laser intensity of $I_{cr} = 2.3\times10^{29}~\trm{Wcm}^{-2}$. An electric field of this order is strong enough to provide a virtual electron-positron pair an energy equal to its rest energy $2mc^2$ in the fleetingly short time $\hbar/mc^2\sim10^{-21}~\trm{s}$ in which the virtual pair ``lives,'' promoting it to reality before the individual particles eventually annihilate with one another. Even at much lower intensities $I$ such as provided by ``strong'' or ultra-intense ($I>10^{23}~\trm{Wcm}^{-2}$) laser fields, the polarised vacuum is predicted to exhibit birefringence and dichroism\cite{dipiazza_PRL_06}, to cause photons to ``merge'' or to ``split'' and even allow them to scatter, all of which awaits experimental confirmation in laser fields. Recent advancements and proposals for the upcoming ELI\cite{ELI_SDR} and HiPER\cite{HiPER_TDR} laser facilities demonstrate a maturing of a technology that will supply intensities of the order $10^{25}\text{-}10^{26}\units{Wcm$^{-2}$}$, which are sufficiently high to test quantum electrodynamics in this relatively unprobed regime.

When driven by a strong electromagnetic field, the virtual electron-positron pairs generate a polarisation and a magnetisation in the vacuum\cite{dipiazza_PRL_06} (see Supplementary Information):
\bea
\mathbf{P}(t,\mbf{r})&=&\phantom{-}\frac{4\alpha^2}{45 m^4}[2(E^{2}-B^{2})\mbf{E} + 7(\mbf{E}\cdot\mbf{B})\mbf{B}]\\ 
\mathbf{M}(t,\mbf{r})&=&-\frac{4\alpha^2}{45 m^4}[2(E^{2}-B^{2})\mbf{B} - 7(\mbf{E}\cdot\mbf{B})\mbf{E}],
\eea
for electric and magnetic fields $\mbf{E}(t,\mbf{r}), \mbf{B}(t,\mbf{r})$. From these expressions, we can form the useful analogy of the polarised vacuum as a solid with non-linear response, which, instead of comprising tangible dipoles, hosts transient polarised virtual particle-antiparticle pairs of dimension approximately equal to the Compton wavelength $\lambdabar_{c}=\hbar/mc\sim10^{-11}~\trm{cm}$. These evanescent pairs mediate a non-linear interaction between fields which becomes more significant the larger the ratio of the applied to the critical field becomes. When discussing strong electromagnetic fields, we are thus referring to a regime totally forbidden in classical physics in which the linear superposition principle in vacuum no more applies.

Taking the solid-state paradigm one step further, using an ultra-intense laser split into two beams, the vacuum can be ``activated'' by polarising two slit-like regions (see Fig. 1). When these regions are probed with a second, (almost) counter-propagating laser, one can imagine creating a real photon-photon double-slit experiment. This employs Babinet's principle, which states that the diffraction pattern of an aperture is identical to that of an opaque obstacle with the same shape as the aperture, justifying our labelling of the two polarised regions as ``slits,'' although they are actually the material-like portion of the scenario. Since accelerated charges radiate, when the polarised vacuum is agitated by the applied field, it forms a source or \emph{vacuum current} of electromagnetic waves, $\mbf{J}_{\trm{vac}}(t,\mbf{r})$. The modified wave equation incorporating vacuum polarisation effects reads (see Supplementary Information):
\be
\nabla^2\mathbf{E}-\frac{1}{c^2}\frac{\partial^2}{\partial t^2}\mathbf{E}=\mbf{J}_{\trm{vac}}(t,\mbf{r}),
\ee
where $\mbf{J}_{\trm{vac}}(t,\mbf{r}) = \protect{(1/c)\nabla\wedge(\partial_{t}\mbf{M}) + (1/c^{2})\partial^{2}_{t}\mbf{P} -  \nabla(\nabla\cdot\mbf{P})}$. This current is then responsible for the generation of two fields $\textbf{E}_{d,i}(t,\mathbf{r})$ with $i=1,2$ each in the centre of the two slits. These fields then interfere to produce the characteristic double-slit diffraction pattern. In Young's original experiment\cite{young04b}, all other incident light was stopped, whereas in our scenario, the probe laser can form a dominant background. Exploiting the wide extension of the field generated in the slits in comparison to the relatively tight focusing of the probe field, allows us to consider regions on the detector plane where the probe background is effectively negligible. This is a significant advancement with respect to previous calculations\cite{dipiazza_PRL_06}, which although based upon the same fundamental physics of quantum electrodynamics and quantum vacuum fluctuations, focused only on the ellipticity and rotation of the polarisation direction acquired by an X-ray probe when it collides with a \emph{single} strong optical standing wave. By introducing the double-slit, the interference pattern of photons generated in the annihilation of virtual electron-positron pairs occurring at different points in space then becomes a useful, measurable quantity and provides in principle both non-local information about the vacuum current and insight on the wave-particle duality of vacuum-generated photons.
 \begin{center}
\begin{figure}
\centering
 \includegraphics[draft=false, width=0.6\linewidth]{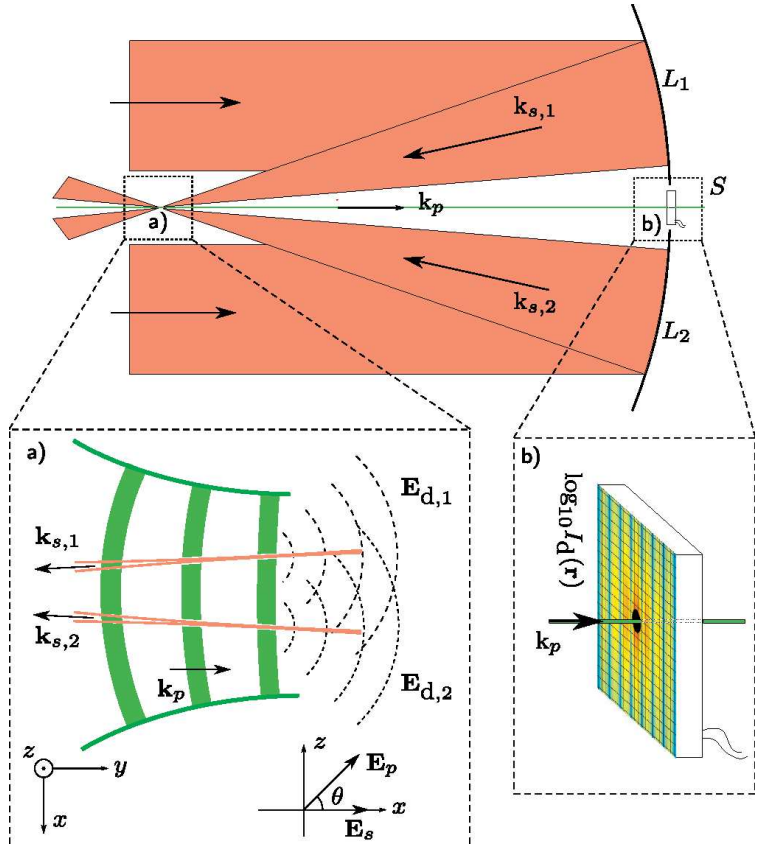}
 \caption{{\bf A matterless double-slit set-up.} Two ultra-intense Gaussian laser pulses with wavevectors $\mbf{k}_{s,1}$ and $\mbf{k}_{s,2}$ are tightly focused by the lenses $L_1$ and $L_2$ (almost) antiparallel to a probe beam with wavevector $\mbf{k}_{p}$ of much wider spot radius (see also inset a)). The vacuum current, activated in the interaction regions of the probe and the strong laser fields, generates photons which interfere to produce a diffraction pattern on the screen $S$. The screen is placed between the focusing mirrors at a distance $y$ along the propagation axis of the probe from the interaction centre and has a hole in the centre allowing the probe beam to pass undisturbed (see also inset b)). The directions of the spatial co-ordinates $x,y,z$ and the angle $\theta$ between the strong field $\mbf{E}_{s}$ and the probe field $\mbf{E}_{p}$ are defined in the inset a).}
\label{exp_setup}
\end{figure}
 \end{center}
With regard to the corresponding experimental implementation, it is pertinent to consider the time-averaged total signal $I_{t}(\mbf{r})$ on a detector plate whose origin is situated in the far field at $\mbf{r} = (0,y,0)$, comprising the scattered fields $\textbf{E}_d(t,\mathbf{r})=\textbf{E}_{d,1}(t,\mathbf{r})+\textbf{E}_{d,2}(t,\mathbf{r})$ and the unperturbed probe field $\mbf{E}_{p}(t,\mbf{r})$: $I_{t}(\mbf{r})=\langle|\mbf{E}_{p}(t,\mbf{r})+\mbf{E}_{d}(t,\mbf{r})|^{2}\rangle=I_{p}(\mbf{r})+I_{pd}(\mbf{r})+I_{d}(\mbf{r})$. Here $I_{p}(\mbf{r})=\langle\mbf{E}_{p}(t,\mbf{r})^{2}\rangle$, $I_{pd}(\mbf{r})=2\langle\mbf{E}_{p}(t,\mbf{r})\cdot\mbf{E}_{d}(t,\mbf{r})\rangle$ and $I_{d}(\mbf{r})=\langle\mbf{E}_{d}(t,\mbf{r})^{2}\rangle$, with $\langle\rangle$ denoting a time average. 

In terms of apparatus, the probe laser should be optical in order that the diffraction pattern is sufficiently large and resolvable. We consider the following nowadays easily-obtainable parameters of 100$\units{fs}$ pulse duration, intensity $4\times10^{16}\units{Wcm$^{-2}$}$ and wavelength $\lambda_{p}=527\units{nm}$ (achievable using the second harmonic of readily-available $1054~\trm{nm}$ lasers with an intensity attenuation of around $2.6$). For the strong-field laser we anticipate parameters available from the upcoming ELI and HiPER facilities of intensity $5\times 10^{24}\units{Wcm$^{-2}$}$, pulse duration $\tau_{s}=30\units{fs}$, wavelength $\lambda_{s}=0.8\units{$\mu$m}$, spot radius $\trm{w}_{s,0}=\lambda_{s}=0.8~\mu\trm{m}$ (corresponding to a laser power of $100\;\text{PW}$) and laser repetition rate $1~\trm{min}$. 

The success of the solid-state perspective is then displayed by plotting the bare vacuum signal $I_{d}(\mbf{r})$ and observing the accuracy of the famous double-slit formula $(n+1/2)\lambda_{p}=D\sin\vartheta$ for predicting minima, indicated by crosses on the $x$-axis in Fig. 2a. $D$ is the distance between the centres of the two ultra-intense lasers, $\vartheta=\tan^{-1}(x/y)$ for detector distance $y$ along the axis from the interaction centre and transverse displacement of the minimum on the detector $x$, with different integers $n$ corresponding to different minima positions. We have chosen to separate the strong beams by $D=80~\trm{w}_{s,0}=64\;\text{$\mu$m}$ and to polarise the probe at $\theta=\pi/2$, focused onto a spot of radius $\trm{w}_{p,0}=290~\mu\trm{m}$ and to place the detector at $y=5\units{m}$. The choice of $D$ is sufficiently large such that the diffraction pattern can be observable also with more realistic (broader) strong beam transverse intensity profiles \cite{Nakatsutsumi_2008}. The single-slit limit of zero strong-field beam separation is depicted in Fig. 2b, in which all fringes have disappeared. The corresponding diffraction pattern does not show typical diffraction rings due to the ``slit'' having edges that are not sharp.

 \begin{center}
\begin{figure}
\centering
 \includegraphics[draft=false, width=0.6\linewidth, clip=true]{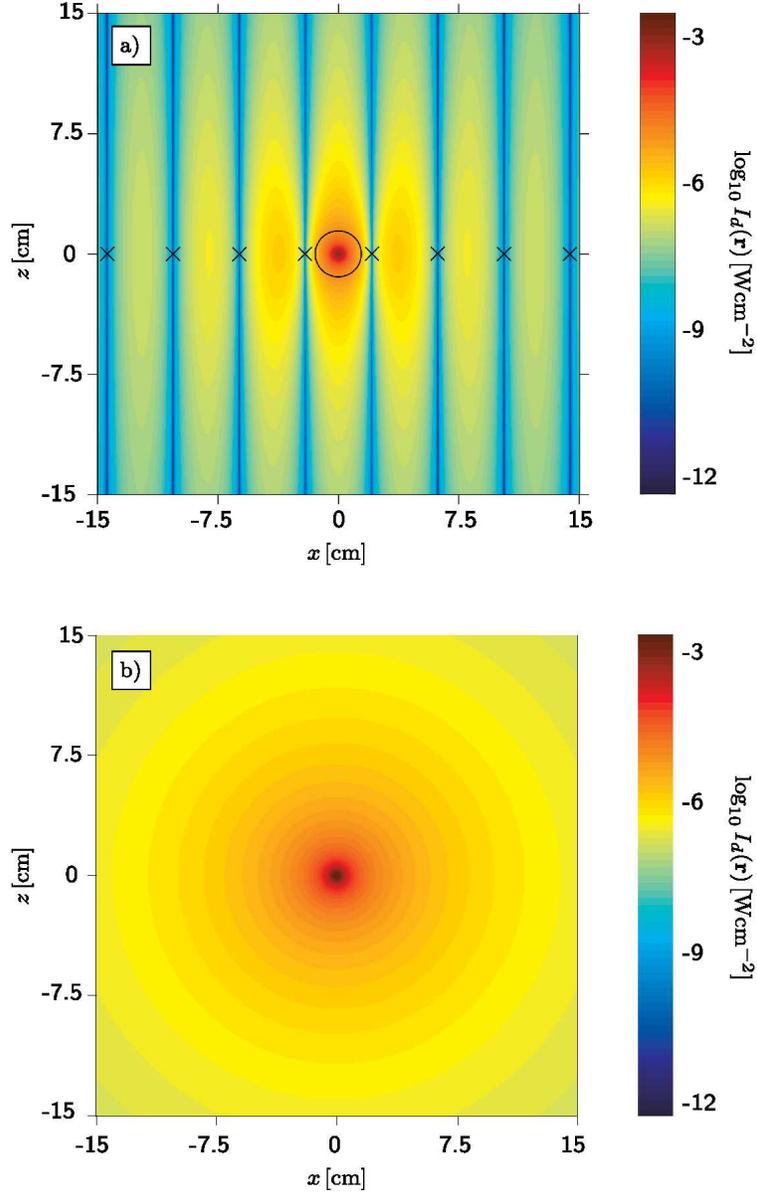}
%
 \caption{{\bf Light-light diffraction pattern.} For the experimental parameters given in the text, the logarithmic plot of the vacuum signal in a) reveals a series of bright and dark fringes resembling the characteristic double-slit pattern. The prediction of the classic $(n+1/2)\lambda_{p}=D\sin\vartheta$ formula for minima is indicated with crosses plotted on the $x$-axis. For a radius greater than $1.5\units{cm}$, indicated on the figure as a black contour, $I_{d}(\mbf{r})$ is much greater than the $I_{p}(\mbf{r})$ and $I_{pd}(\mbf{r})$ background. For the same parameters as a) but at zero strong-beam separation, the logarithmic plot of the vacuum signal in the far-field detector plane is shown in b).}
\label{diff_reses}
\end{figure}
\end{center}
Turning to quantitative results, we envisage verifying the phenomenon by either a full or partial resolution of the interference pattern, or else by simple counting of diffracted photons. At our probe's wavelength of $527~\trm{nm}$ back-illuminated CCDs (charge-coupled devices) have an efficiency of $90\%$ \cite{ccd}. From numerical results for the aforementioned typical experimental parameters, on a region in which $I_{d}(\mbf{r})$ is more than one-hundred times larger than $I_{p}(\mbf{r})$ and $I_{pd}(\mbf{r})$, taking into account CCD efficiency, we expect per shot of the strong field, $4$ photons from the vacuum signal. By repeating the experiment first in the absence of the strong beams and then the probe and vice versa, one is in principle able to account for possible background photons coming from those beams. Background photons with a frequency different to that of the probe could be excluded by placing frequency filters in front of the detector screen. The presence of a thermal photon background can then be neglected operating at a typical temperature of the order of $300\;\text{K}$. Moreover, a good vacuum quality of the order of $10^{-6}\text{-}10^{-5}\;\text{torr}$ is required in order to neglect diffraction effects due to the presence of residual gas in the interaction region. We have also ensured that with the above numerical parameters, alterations to the vacuum signal due to the pulse shape of the strong beams can be consistently neglected.

One can form the visibility of the diffraction pattern (see \cite{scully97} and Supplementary Information), to determine how many photons are required before fringes can be adequately differentiated. For a scenario in which a $15\units{cm}\times15\units{cm}$ CCD with a central circular aperture of $1.5\;\text{cm}$ radius, placed as indicated in Fig. 1, detects vacuum signal photons for the aforementioned experimental parameters, a theoretical maximum visibility of $47.6\%$ can be reached. After modelling experimental trials numerically, it was found that $\sim$ 1000 photons were required before the statistical fluctuations around this analytical value were reduced to less than $10\%$, corresponding to an operating time of approximately four hours (see Supplementary Information).

By exploiting the polarised vacuum in such a scenario with lasers available in the next few years, one can take Young's famous experiment one step further and create a truly quantum double-slit set-up comprising entirely of light. In addition, by counting photons or measuring the intensity pattern directly, such a method can be employed to probe the quantum vacuum and to study its structure as predicted by quantum electrodynamics.

\clearpage

\section*{Supplementary Information}

In the limit of electromagnetic fields $\mathbf{E}(t,\mathbf{r})$ and $\mathbf{B}(t,\mathbf{r})$ with amplitude much less than the critical fields $E_{\text{cr}}=\sqrt{4\pi}m^2c^3/\hbar e=1.3\times 10^{16}\;\text{V/cm}$ and $B_{\text{cr}}=\sqrt{4\pi}m^2c^3/\hbar e=4.4\times 10^{13}\;\text{G}$ (in our units the fine-structure contant reads $\alpha=e^2/4\pi\hbar c\approx 1/137$) and with wavelength much larger than the Compton wavelength $\lambdabar_{\trm{c}}=\hbar/mc=3.9\times 10^{-11}\;\text{cm}$ the vacuum Lagrangian density of the electromagnetic field including quantum correcting terms due to vacuum polarization is given by\cite{heisenberg_euler36}:
\be
\mathcal{L} = \frac{1}{2} (E^{2} - B^{2}) + \frac{2\alpha^{2}}{45m^{4}} 
              \big[ (E^{2} - B^{2})^{2} + 7(\mathbf{E} \cdot \mathbf{B})^{2} \big],\label{eqn:EH-L}
\ee
where units $\hbar=c=1$ are employed and terms proportional to $\alpha^{2}$ represent quantum corrections much smaller than the Maxwell Lagrangian density $(E^2-B^2)/2$. In our scenario, the total electromagnetic field consists of two strong Gaussian-focused beams linearly polarized along the $x$-direction that propagate along the negative $y$-direction antiparallel to a weaker Gaussian-focused probe beam linearly polarized at an angle $\theta$ to the $x$-axis. The strong beams have frequency $\omega_{\text{s}}$ (wavelength $\lambda_{\text{s}}=2\pi/\omega_{\text{s}}$), peak electric field $E_{\text{s},0}/\sqrt{2}$ and are centred at $(x,z) = \pm(x_{0},z_{0})$ with waists $w_{\text{s}}(y)=w_{\trm{s},0}\sqrt{1+(y/y_{\trm{r},\trm{s}})^{2}}$ and Rayleigh length $y_{\trm{r},\trm{s}}=\omega_{\trm{s}}w_{\trm{s},0}^{2}/2$. The probe beam has frequency $\omega_{\trm{p}}$ (wavelength $\lambda_{\trm{p}}=2\pi/\omega_{\trm{p}}$), peak electric field $E_{\trm{p},0}$ and is centred at $(x,z) = (0,0)$ with waist $w_{\trm{p}}(y)=w_{\trm{p},0}\sqrt{1+(y/y_{\trm{r},\trm{p}})^{2}}$, $w_{\trm{p},0}\gg w_{\trm{s},0}$ and Rayleigh length $y_{\trm{r},\trm{p}}=\omega_{\trm{p}}w_{\trm{p},0}^{2}/2$. Here we concentrate on the currently unknown leading order contribution of elastic real photon-photon scattering and in our analysis correspondingly take a solution of Maxwell's equations to first order in $(w_{j,0}/y_{\trm{r},j})$ with $j=\trm{s},\trm{p}$ (see e. g. \cite{salamin_review06} for more details on the approximation used here). We also neglect the angle between the strong laser beams (see Fig. 1 in the main text) and assume they propagate in the same direction. Angles of the order of $0.1\text{-}0.2\;\text{rad}$ can be in principle achieved experimentally and, following our numerical simulations, lead to corrections of the order of $10\text{-}20\;\%$, respectively. The following equations are used to represent the system:
\bea
\mbf{E}(t,\mbf{r}) & = & \mbf{E}_{\trm{s}}(t,\mbf{r}) + \mbf{E}_{\trm{p}}(t,\mbf{r})\\
\mbf{E}_{\trm{s}}(t,\mbf{r}) &:=& [E_{\trm{s},1}(t,\mbf{r}) + E_{\trm{s},2}(t,\mbf{r})]\mbf{\hat{x}}\\
E_{s,1/2}(t,\mbf{r}) & := & \frac{1}{\sqrt{2}}\mathcal{E}_{\trm{s},0}(x\mp x_{0},y,z\mp z_{0}) \sin \Big(\omega_{\trm{s}}(t + y) - f_{\trm{s}}(x\mp x_{0},y,z\mp z_{0})\Big)\nonumber\\
\mbf{E}_{\trm{p}}(t,\mbf{r}) & := & E_{\trm{p}}(t,\mbf{r}) [\mbf{\hat{x}}\cos\theta+\mbf{\hat{z}}\sin\theta] \\
E_{\trm{p}}(t,\mbf{r}) & := & \mathcal{E}_{\trm{p},0}(x,y,z) \sin \Big(\omega_{\trm{p}}(t - y) + f_{\trm{p}}(x,y,z)\Big),\nonumber
\eea
where
\be
\mathcal{E}_{j,0}(x,y,z) := \frac{E_{j,0}e^{-(x^{2} + z^{2})/w_{j}^{2}(y)}} 
                                          {\sqrt{1+(y/y_{\trm{r},j})^{2}}},
\ee
and
\be
f_{j}(x,y,z) = \psi_{j} + \tan^{-1}\Big(\frac{y}{y_{\trm{r},j}}\Big) - \frac{\omega_{j}y}{2} \frac{x^2 + z^{2}}{y^2 + y_{\trm{r},j}^{2}},
\ee
where $\psi_{j}$ is a constant phase offset, $j=\trm{s},\trm{p}$. The inclusion here of defocusing terms in the probe field $\mbf{E}_{\trm{p}}(t,\mbf{r})$ scaling as $y/y_{\trm{r},\trm{p}}$ is a significant improvement on previous results\cite{dipiazza_PRL_06}, allowing us to investigate the vacuum polarization effects also in the so-called far region, where the observation distance $y$ is so large that $y/y_{\trm{r},\trm{p}}\gg 1$. This is essential here, as the suggested experimental setup requires the observation screen to be located far from the interaction region (in the numerical example considered in the main text we have $y/y_{\trm{r},\trm{p}}\approx 10$).

By applying the principle of least action to the Lagrangian density in Eq. (\ref{eqn:EH-L}), one obtains the wave equation for the total electric field:
\bea\label{Wave_Eq}
\nabla^{2} \mbf{E} - \partial_{t}^{2} \mbf{E} & = & \mathbf{J}_{\trm{vac}},
\eea
where the ``vacuum current'' $\mathbf{J}_{\trm{vac}}(t,\mathbf{r})$ can be written as
\bea
\mathbf{J}_{\trm{vac}}=\nabla \wedge \partial_{t}\mbf{M}
                                                     -\nabla (\nabla \cdot \mbf{P})
                                                     +\partial_{t}^{2}\mbf{P},
\eea
with $\mbf{P}(t,\mathbf{r})$ and $\mbf{M}(t,\mathbf{r})$ being the vacuum polarization and magnetization respectively:
\bea
\mbf{P} & := & \phantom{-}\frac{4 \alpha^{2}}{45 m^4} \big[ 2(E^{2} - B^{2}) \mbf{B} 
                                             + 7(\mbf{E} \cdot \mbf{B}) \mbf{B} \big] \\
\mbf{M} & := & -\frac{4 \alpha^{2}}{45 m^4} \big[ 2(E^{2} - B^{2}) \mbf{B} 
                                          - 7(\mbf{E} \cdot \mbf{B}) \mbf{E} \big].
\eea
The wave equation (\ref{Wave_Eq}) can be written formally as the integral equation
\be
\mbf{E}(t,\mbf{r}) =\mbf{E}_{\trm{cl}}(t,\mbf{r})+ \int dt'd^3\mathbf{r}'D(t-t',\mathbf{r}-\mathbf{r}')\mbf{J}_{\trm{vac}}(t',\mathbf{r}')\label{eqn:Ed_def},
\ee
by employing the Green's function $D(t,\mathbf{r})=-1/(2\pi)^{4}\int d\omega d\mathbf{k}\exp[-i(\omega t-\mathbf{k}\cdot \mathbf{r})]/(\omega^2-|\mathbf{k}|^2)$ (see, for example, \cite{jackson99}). The first term in this equation is the classical solution that in our case is given by $\mbf{E}_{\trm{cl}}(t,\mbf{r})=\mbf{E}_{\trm{s}}(t,\mbf{r}) + \mbf{E}_{\trm{p}}(t,\mbf{r})$. The second term arises due to the quantum interaction between the probe and the strong fields, which we label the diffracted field $\mbf{E}_{\trm{d}}(t,\mbf{r})$, and is calculated by substituting the zero-order solution $\mbf{E}_{\trm{cl}}(t,\mbf{r})$ into the vacuum current $\mathbf{J}_{\trm{vac}}(t,\mathbf{r})$. Since our probe and strong fields are monochromatic it is convenient to work in the Fourier-transformed frequency space and the diffracted field can be written as:
\be\label{FF}
\mbf{E}_{\trm{d}}(t, \mbf{r}) = \mbf{E}_{\trm{d}}(\mbf{r})\frac{e^{i(\omega_{\trm{p}}(t-r)+\psi_{\trm{p}})}}{2i} + \trm{c.c.}\;.
\ee
The Fourier amplitude $\mbf{E}_{\trm{d}}(\mbf{r})$ is then given by
\be
\mbf{E}_{\trm{d}}(\mbf{r}) = \frac{I_{\trm{s},0}}{I_{\trm{cr}}} \frac{\alpha E_{\trm{p}}\mbf{v}}{90}\sum_{k=1}^{4}\mathcal{I}_{\trm{k}}(\mbf{r})\label{Ed_final},
\ee
where $I_{\trm{s},0}=E_{\trm{s},0}^2/2$ is the strong field intensity, $\mbf{v}$ is the polarization vector with:
\be
\mbf{v} = \left(
\begin{array}{c}
4(1+\frac{y}{r})\cos\theta\\
-(1-\frac{y}{r})(4 \frac{x}{r}\cos\theta +7\frac{z}{r}\sin\theta)\\
7(1+\frac{y}{r})\sin\theta\\
\end{array}
\right) + \trm{O}\left(\Big(\frac{x}{r}\Big)^{2}, \Big(\frac{z}{r}\Big)^{2} \right), \label{v_vec}
\ee
and $\mathcal{I}_{\trm{k}}(\mbf{r})$ are the four integration volumes:
\bea
\mathcal{I}_{\trm{k}}(\mbf{r}) &=& \int^{\infty}_{-\infty}\,d^{3}r' \frac{e^{\mathscr{F}_{\trm{k}}}}{(1+(y'/y_{\trm{r},\trm{s}})^{2})\sqrt{1+(y'/y_{\trm{r},\trm{p}})^{2}}}\\
\mathscr{F}_{\trm{k}} &=& -i\omega_{\trm{p}}\Big(y' + \frac{x'^{2}+y'^{2}+z'^{2}}{2r} - \frac{xx'+yy' +zz'}{r} \nonumber \\
&&- \frac{(xx'+yy'+zz')^{2}}{2r}\Big) - \frac{2}{w_{\trm{s}}^{2}(y')}(x'^{2}+z'^{2}+x_{0}^{2}+z_{0}^{2})\nonumber\\
&&-\frac{x'^{2}+z'^{2}}{w_{\trm{p}}^{2}(y')}+i\tan^{-1}\frac{y'}{y_{\trm{r},\trm{p}}}-\frac{i\omega_{\trm{p}}y'}{2}\frac{x'^{2}+z'^{2}}{y'^{2}+y_{\trm{r},\trm{p}}^{2}} \nonumber\\
&& + (x'x_{0}+z'z_{0}) \Big(\frac{4\beta_{\trm{k}}}{w_{\trm{s}}^{2}(y')} + \frac{2i\Gamma_{\trm{k}}\omega_{\trm{s}}y'}{y'^{2}+y_{\trm{r},\trm{s}}^{2}}\Big) + i\Gamma_{\trm{k}}\Delta\psi_{\trm{s}},
\eea
where ${\beta}_{1} = 1, {\beta}_{2} = -1$ and ${\beta}_{3}={\beta}_{4}=0$, $\Gamma_{1} = \Gamma_{2} = 0$, $\Gamma_{3}=1$ and $\Gamma_{4}=-1$, where $\Delta\psi_{\trm{s}} = \psi_{\trm{s},2}-\psi_{\trm{s},1}$. $|\mbf{v}\cdot\mbf{v}|=\trm{v}^{2}$ is maximized for $\cos2\theta=-1$, and without loss of generality, we set $\Delta\psi_{\trm{s}}$ and $z_{0}$ to zero.

Since the diffracted field contains spatial integrals over the probe and the strong fields, its decay length in the transverse $x$-$z$ plane results in being much larger than that of the probe field. It can be shown that the integrals $\mathcal{I}_{3}(\mbf{r})$ and $\mathcal{I}_{4}(\mbf{r})$ are negligible with respect to $\mathcal{I}_{1}(\mbf{r})$ and $\mathcal{I}_{2}(\mbf{r})$, which depend only on the physical parameters of one of the strong beams respectively and therefore describe the interaction of the probe field with each ``slit.'' Therefore the diffracted field amplitude $\mbf{E}_{\trm{d}}(\mbf{r})$ can be written as $\mbf{E}_{\trm{d}}(\mbf{r}) = \mbf{E}_{\trm{d},1}(\mbf{r})+\mbf{E}_{\trm{d},2}(\mbf{r})$ with the subscripts $1,2$ referring to the respective terms in \eqnref{Ed_final} and the quantities $\mbf{E}_{\trm{d},i}(t,\mbf{r})$ ($i=1,2$) employed in the text derived from Eq. (\ref{FF}).

The analytical expression for $I_{\trm{d}}(\mbf{r})=\langle\mbf{E}_{\trm{d}}(t,\mbf{r})^{2}\rangle$, with $\langle\rangle$ denoting a time average, in the limit of no probe focusing, $y_{\trm{r},\trm{p}}\rightarrow\infty$, $w_{\trm{p},0}\gg w_{\trm{s},0}$ and $x/r, z/r \ll y/r\approx 1$ and zero beam separation $x_{0}=z_{0}=0$ was also derived and found to have excellent agreement with numerical results. As a further check, we derived the ellipticity $\varepsilon$ induced in the probe and compare this to the expression for two, parallel-propagating, colliding lasers in the \emph{refractive-index} (i. e. short observation distances, $y\rightarrow 0$), \emph{crossed-field} ($\omega_{\trm{s}}\rightarrow0$) limit found in other literature\cite{heinzl_birefringence06}. One achieves the result $\eps(\theta=\pi/4)=(2\alpha\pi/15)(I_{\trm{s},0}/I_{\trm{cr}})(l_{\trm{y}}/\lambda_{\trm{p}})$, where $l_{\trm{y}}$ is the effective interaction length of the two sets of beams, $l_{\trm{y}}=\pi y_{\trm{r},\trm{p}}y_{\trm{r},\trm{s}}/(y_{\trm{r},\trm{p}}+y_{\trm{r},\trm{s}})$, which then agrees in the limit $y_{\trm{r},\trm{p}}\rightarrow\infty$ with the literature\cite{heinzl_birefringence06}.

For a fixed $y$, we can approximately maximize the region in which $I_{\trm{d}}(\mbf{r})\gg I_{\trm{p}}(\mbf{r})+I_{\trm{pd}}(\mbf{r})$, with $I_{\trm{p}}(\mbf{r})=\langle\mbf{E}_{\trm{p}}(t,\mbf{r})^2\rangle$ and $I_{\trm{pd}}(\mbf{r})=2\langle\mbf{E}_{\trm{p}}(t,\mbf{r})\cdot\mbf{E}_{\trm{d}}(t,\mbf{r})\rangle$, and hence maximize the signal-to-noise ratio by ensuring the fastest decay of the probe and mix-term background in this far-field plane. This occurs when the Gaussian variance is minimized, corresponding to a probe focused to a waist of $w'_{\trm{p},0}=\sqrt{\lambda_{\trm{p}}y/2\pi}$.

The results of the numerical photon-counting experiments are given in Fig. 1. For segments of anticipated maxima and minima of intensity of the same width, $I_{\trm{max}}$, $I_{\trm{min}}$, in a region of the detector plate where $I_{\trm{d}}(\mbf{r})$ is much larger than the background, the visibility $V$ is given by: $V=(I_{\trm{max}}-I_{\trm{min}})/(I_{\trm{max}}+I_{\trm{min}})$. Each experiment consisted of generating photons randomly on the detector plate according to a probability distribution given by the numerical solution to the diffracted field intensity. Only the region in which $I_{\trm{d}}(\mbf{r}) >100[I_{\trm{p}}(\mbf{r})+I_{\trm{pd}}(\mbf{r})]$ was retained in $I_{\trm{d}}(\mbf{r})$, which was then summed in the $z$-direction, a direction of approximate symmetry, and subsequently normalized to generate the one-dimensional probability density function used in each numerical trial. Numerical modelling of 10,000 successive experimental trials was performed, in which the visibility was repeatedly calculated for each new incident photon in the trial, with a total of 10,000 photons per trial. This allowed us to determine how the visibility fluctuated around the analytical value of $V_{0}=47.6\%$, where the fluctuations in general decreased with more detected photons. This produced stochastic trails for each trial, in which the largest number of photons where the fluctuation was greater than a given value (indicated on the horizontal axis of Fig. 1) was taken to be the number of photons required to reach that accuracy in the visibility.
 \begin{figure}
 \begin{center}
 \includegraphics[draft=false, width=0.9\linewidth]{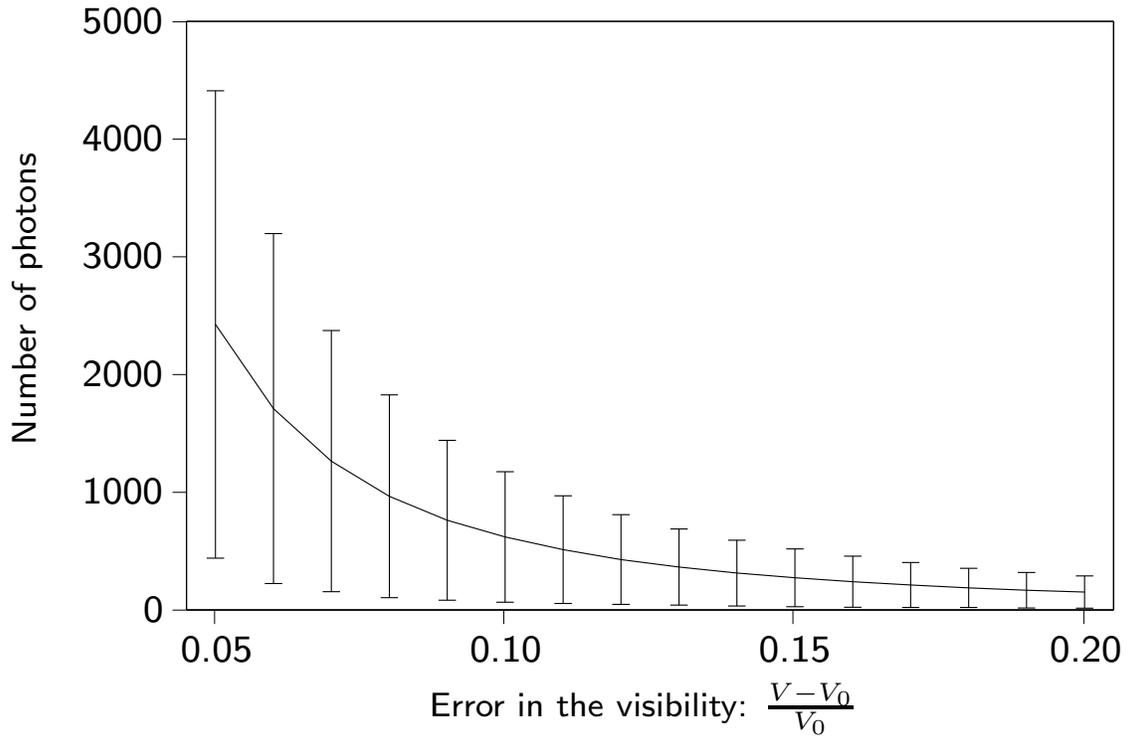}
 \end{center}
 \caption{{\bf Counting photons.} The number of photons required to reduce the error in the visibility below the value given on the horizontal axis is plotted after averaging results of 10,000 experimental trials, each with 10,000 photons, with the error bars given by the standard deviation over the trials.}\label{fig:pcount}
 \end{figure}

%
%

\bibliography{standard_nature}

\end{document}